\documentclass[twocolumn,showpacs,preprintnumbers,amsmath,amssymb,superscriptaddress]{revtex4-1}

\usepackage[pdftex]{graphicx}
\usepackage{amsmath}
\allowdisplaybreaks

\begin{document}


\title{Evidence for an extended critical fluctuation region above the polar ordering transition in LiOsO$_3$}


\author{Jun-Yi Shan}
\affiliation{Department of Physics, California Institute of Technology, Pasadena, California 91125, USA\looseness=-1}
\affiliation{Institute for Quantum Information and Matter, California Institute of Technology, Pasadena, California 91125, USA\looseness=-1}
\author{A. \surname{de la Torre}}
\affiliation{Department of Physics, California Institute of Technology, Pasadena, California 91125, USA\looseness=-1}
\affiliation{Institute for Quantum Information and Matter, California Institute of Technology, Pasadena, California 91125, USA\looseness=-1}
\author{N. J. Laurita}
\affiliation{Department of Physics, California Institute of Technology, Pasadena, California 91125, USA\looseness=-1}
\affiliation{Institute for Quantum Information and Matter, California Institute of Technology, Pasadena, California 91125, USA\looseness=-1}
\author{L. Zhao}
\affiliation{Department of Physics, University of Michigan, Ann Arbor, Michigan 48109, USA\looseness=-1}
\author{C. D. Dashwood}
\affiliation{London Centre for Nanotechnology and Department of Physics and Astronomy, University College London, London WC1E 6BT, UK\looseness=-1}
\author{D. Puggioni}
\affiliation{Department of Materials Science and Engineering, Northwestern University, Illinois 60208-3108, USA\looseness=-1}
\author{C. X. Wang}
\affiliation{Institute of Physics, Chinese Academy of Sciences, Beijing 100190, China\looseness=-1}
\author{K. Yamaura}
\affiliation{Research Center for Functional Materials, National Institute for Materials Science, 1-1 Namiki, Tsukuba, Ibaraki 305-0044, Japan\looseness=-1}
\author{Y. Shi}
\affiliation{Institute of Physics, Chinese Academy of Sciences, Beijing 100190, China\looseness=-1}
\author{J. M. Rondinelli}
\affiliation{Department of Materials Science and Engineering, Northwestern University, Illinois 60208-3108, USA\looseness=-1}
\author{D. Hsieh}
\email[Corresponding author: ]{dhsieh@caltech.edu}
\affiliation{Department of Physics, California Institute of Technology, Pasadena, California 91125, USA\looseness=-1}\affiliation{Institute for Quantum Information and Matter, California Institute of Technology, Pasadena, California 91125, USA\looseness=-1}

\date{\today}

\begin{abstract}

Metallic LiOsO$_3$ undergoes a continuous ferroelectric-like structural phase transition below $T_c$ = 140 K to realize a polar metal. To understand the microscopic interactions that drive this transition, we study its critical behavior above $T_c$ via electromechanical coupling --- distortions of the lattice induced by short-range dipole-dipole correlations arising from Li off-center displacements. By mapping the full angular distribution of second harmonic electric-quadrupole radiation from LiOsO$_3$ and performing a simplified hyper-polarizable bond model analysis, we uncover subtle symmetry-preserving lattice distortions over a broad temperature range extending from $T_c$ up to around 230 K, characterized by nonuniform changes in the short and long Li-O bond lengths. Such an extended region of critical fluctuations may explain anomalous features reported in specific heat and Raman scattering data, and suggests the presence of competing interactions that are not accounted for in existing theoretical treatments. More broadly, our results showcase how electromechanical effects serve as a probe of critical behavior near inversion symmetry breaking transitions in metals.
\end{abstract}

\maketitle

\section{Introduction}
Ferroelectric phase transitions typically occur in insulating materials where long-range electrostatic forces between local electric dipoles are unscreened \cite{Slater}. The observation of a ferroelectric-like structural phase transition in metallic LiOsO$_3$ in 2013 \cite{Shiferroelectriclikestructuraltransition2013a} was therefore counterintuitive and challenged the conventional understanding of how polar distortions are stabilized \cite{AndersonSymmetryConsiderationsMartensitic1965b,Puggioni2014}. Over the last few years, there is growing evidence that the phase transition in LiOsO$_3$ is primarily of a continuous order-disorder type \cite{Shiferroelectriclikestructuraltransition2013a,SimFirstprinciplesstudyoctahedral2014a,LiuMetallicferroelectricityinduced2015a,JinRamanphononsferroelectriclike2016b,PadmanabhanLinearnonlinearoptical2018a,JinRamaninterrogationferroelectric2019}, where local dipole moments generated by the off-center displacement of Li ions form well above the Curie temperature ($T_c \sim$ 140 K). Based on first-principles calculations \cite{XiangOriginpolardistortion2014a,BenedekFerroelectricmetalsreexamined2016,ZhaoMetascreeningpermanencepolar2018a}, these local moments interact through short-range forces arising from local bonding preferences, which are un-screened by the itinerant carriers, to achieve long-range order at $T_c$.

To understand the form of the short-range interacting Hamiltonian that describes LiOsO$_3$, it is necessary to address its critical behavior upon cooling through $T_c$. In insulating proper ferroelectrics such as the isostructural compound LiNbO$_3$, the presence of long-range forces should suppress dipolar fluctuations and lead to a very narrow or no critical region \cite{Tokunaga1974,Tokunaga1976,Ginzburg}. In polar metals, on the other hand, where inversion symmetry is lifted through geometric routes, the screening of long-range forces can in principle lead to wide critical regions that are more amenable to study. However, critical dipolar fluctuations in LiOsO$_3$ cannot be probed using standard dielectric measurements owing to its metallicity. They are also challenging to measure using x-ray or neutron scattering due to the weak interaction with Li and the small size of available single crystals respectively. The temperature dependence of the local dielectric susceptibility of LiOsO$_3$ was recently inferred from the linewidth of a Raman active Li vibrational mode \cite{JinRamaninterrogationferroelectric2019}, but it was argued to closely follow a Curie-Weiss law above $T_c$, with no evidence of a critical region.

\begin{figure}[t]
\includegraphics[width=1\columnwidth]{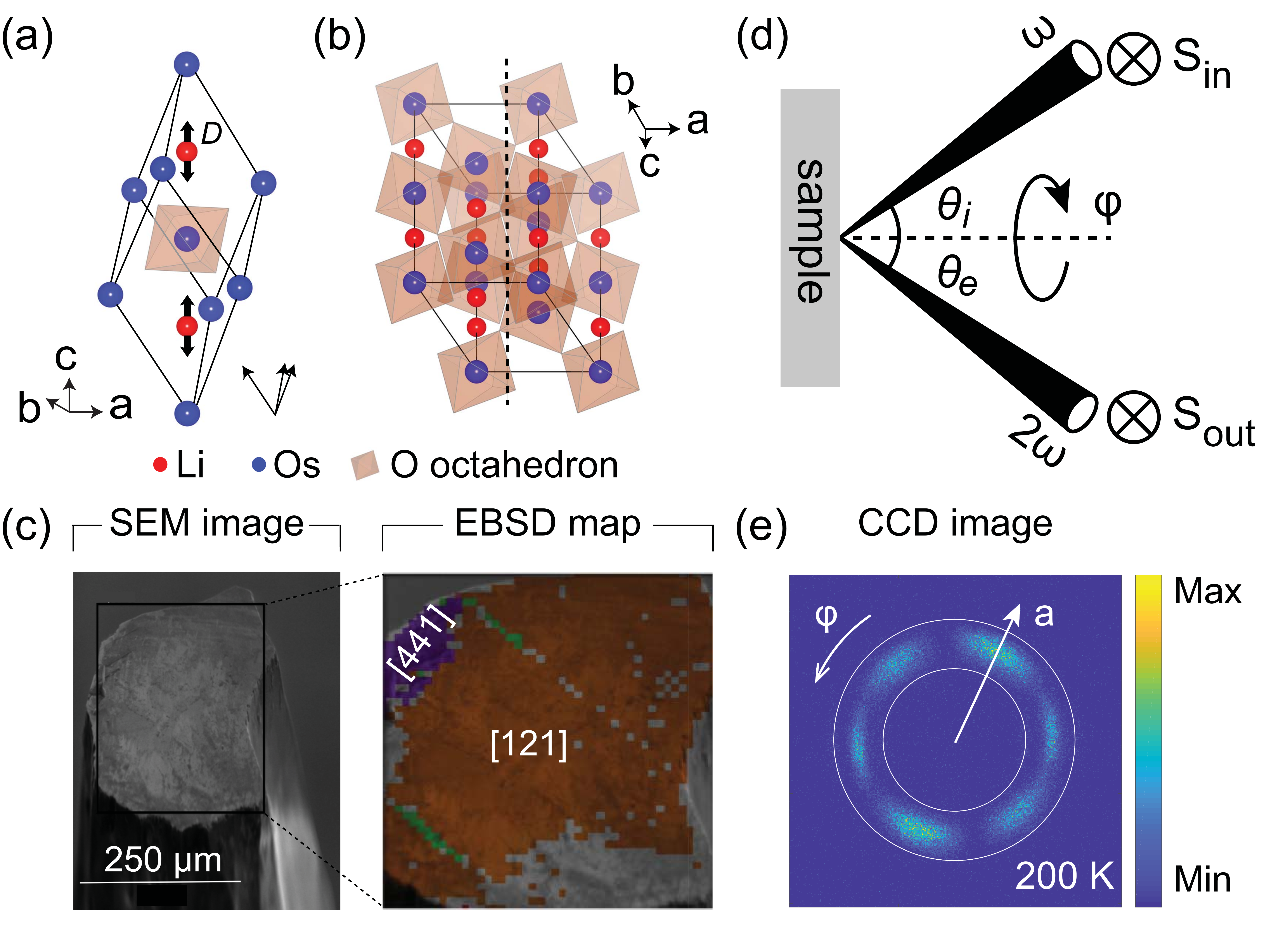}
\caption{\label{fig1} (a) Crystal structure of LiOsO$_{3}$ in the nonpolar $R\overline{3}c$ phase. Black arrows indicate the two possible polar displacements of the Li atoms that produce the dipole moments $D$ in the polar $R3c$ phase. Crystallographic axes on the bottom left (right) are in the hexagonal (rhombohedral) setting. (b) Projection of the crystal structure onto the studied [121] surface. The black dashed line is the projection of the polar $c$ axis onto this surface. (c) Scanning electron microscopy (SEM) image and enlarged electron backscatter diffraction (EBSD) map of a typical polished [121] surface. Different colors on the EBSD map represent different crystal orientations. (d) Schematic of the SHG-RA experiment: laser light with photon energy $\hbar\omega$ = 1.5 eV is focused onto the sample at some chosen angle of incidence $\theta_i$. The emitted SHG light at 3 eV is collected as a function of the scattering plane angle $\varphi$ and the angle of emission $\theta_e$ using a CCD camera. All data shown in the main text were acquired in the S$\mathrm{_{in}}$-S$\mathrm{_{out}}$ polarization geometry, but data in other geometries provide consistent results \cite{EPAPS}. (e) A typical raw SHG-RA data set from LiOsO$_3$ measured above $T_c$ for a fixed $\theta_i$. The $\varphi$ and $\theta_e$ dependence are projected respectively along the azimuthal and radial directions in the CCD image. The white arrow is the crystallographic $a$ axis.}
\end{figure}

In this article, we detect short-range dipolar correlations arising from Li cation displacements above $T_c$ in LiOsO$_3$ through electromechanical coupling, which is normally seen in insulators. In polar metals, this effect can also be large due to microscopic dipole-strain interactions. More specifically, the short-range interaction energy between local $c$-axis oriented moments $D_{i}$ in LiOsO$_3$ [Fig.~\ref{fig1}(a)] can be described by the thermal expectation value of an Ising-type Hamiltonian $\mathcal{H}_{int}=\sum\limits_{\langle ij\rangle}J_{ij}D_{i}D_{j}$ \cite{BarretoFerrroelectricphasetransitions2000a,LiuMetallicferroelectricityinduced2015a,JinRamaninterrogationferroelectric2019}, where the coupling constants $J_{ij}$ depend sensitively on the position of atoms in the vicinity of sites $i$ and $j$. The atomic positions minimize the total elastic energy of the system $ E_T = \langle\mathcal{H}_{int}\rangle + \langle\mathcal{H'}\rangle$, where $\mathcal{H'}$ includes all other interactions. As the dipole correlators $\langle D_{i}D_{j}\rangle$ increase upon cooling, atomic positions will shift to readjust the balance between $\langle\mathcal{H}_{int}\rangle$ and $\langle\mathcal{H'}\rangle$ in order to minimize $E_T$. Therefore, $\langle D_{i}D_{j}\rangle$ can be tracked via subtle symmetry-preserving changes in atomic coordinates. By detecting such electrostrictive effects in LiOsO$_3$ using high-multipole optical second harmonic generation rotational anisotropy (SHG-RA), we reveal a wide critical region extending from $T_c$ up to $T' \sim$ 230 K, the characteristic temperature where short-range correlations start to grow. The observation of strong fluctuations suggests the presence of competing short-range interactions in the system, which is not captured by existing density functional theory or effective model calculations \cite{LiuMetallicferroelectricityinduced2015a}, and is consistent with specific heat data reporting a low entropy loss across $T_c$ \cite{Shiferroelectriclikestructuraltransition2013a}.

\begin{figure}[t]
\includegraphics[width=0.95\columnwidth]{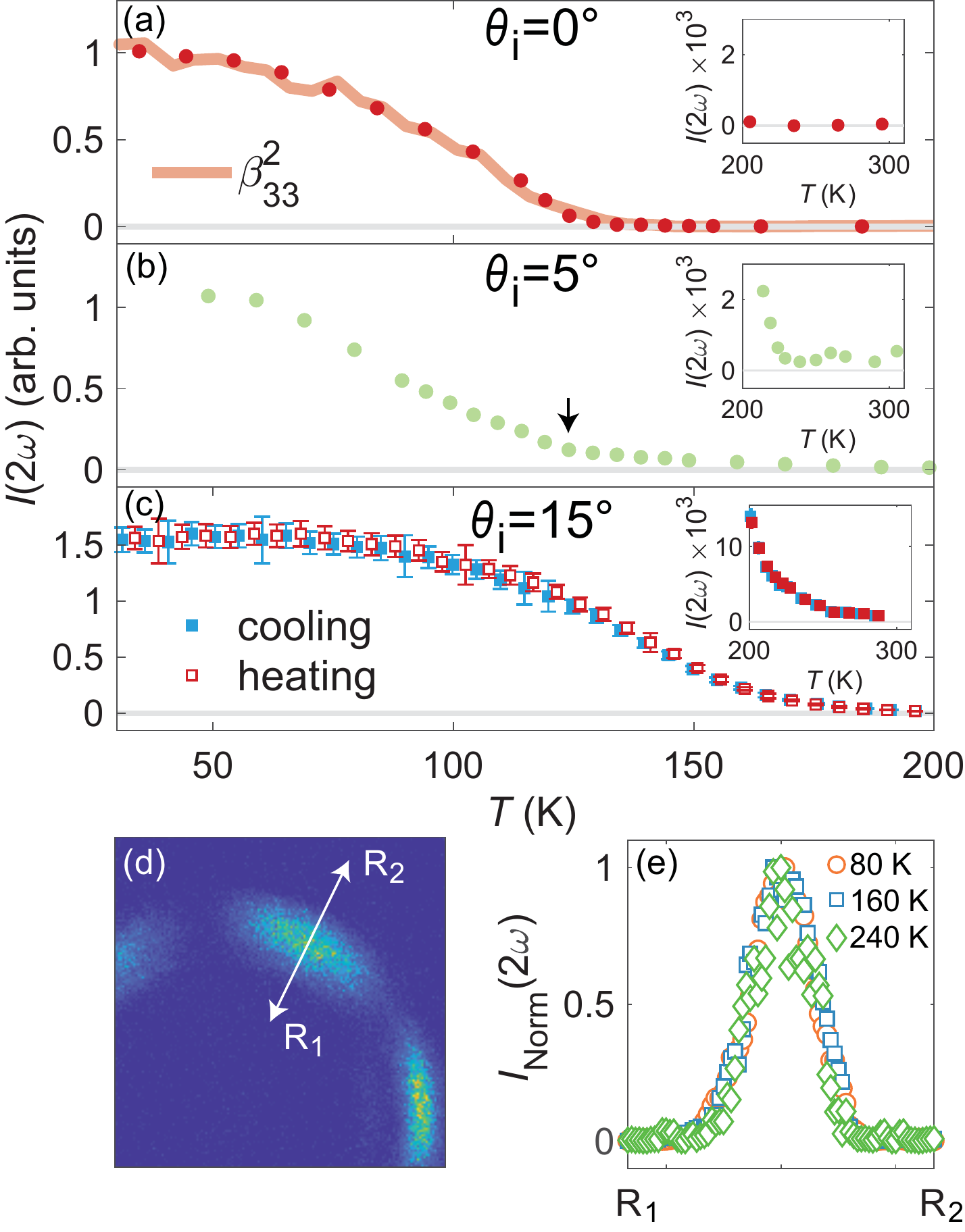}
\caption{\label{fig2} (a)-(c) Temperature-dependent SHG intensity from LiOsO$_3$ at fixed $\varphi$ and $\theta_{i}$ = 0$^{\circ}$, 5$^{\circ}$ and 15$^{\circ}$, all scaled to the intensity of the $\theta_{i}$ = 0$^{\circ}$ curve at base temperature. The intensities are integrated over the $\theta_e$ range bounded by the concentric white circles shown in Fig.~\ref{fig1}(e). Insets show a close-up on the temperature range between 200 K and 300 K. The square of the Li displacement parameter $\beta_{33}$ reported in a neutron diffraction study \cite{Shiferroelectriclikestructuraltransition2013a} is overlaid on the normal incidence data (panel a). We shifted the reported $\beta_{33}$ curve downwards by 8 K to account for laser heating of the sample in our experiments. The black arrow (b) indicates the kink mentioned in the main text. No difference between cooling and heating curves was observed. (d) The $\theta_e$ dependence of the SHG intensity is extracted from a cut along the radial direction of the CCD image. (e) The normalized intensity profile along a radial cut from R$_1$ to R$_2$ taken at different temperatures.}
\end{figure}

\begin{figure*}[t]
\includegraphics[width=2\columnwidth]{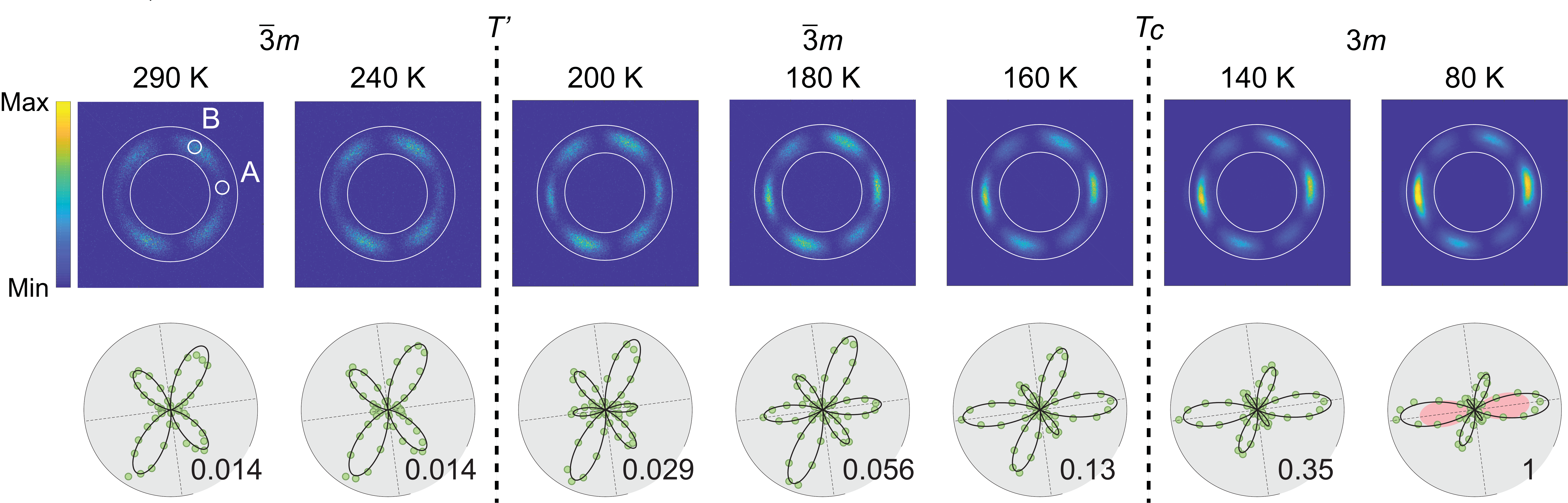}
\caption{\label{fig3} The top row shows raw SHG-RA data sets from LiOsO$_3$ at select temperatures taken with a fixed $\theta_{i} = 15^{\circ}$. The bottom row shows polar plots of the data (green circles) obtained by radially integrating the raw data between the concentric white circles shown in the top row. Fits to the simplified hyperpolarizable bond model are superposed (black lines). For the 80 K fit we include an ED SHG contribution, with a functional form based on the experimental normal incidence RA pattern (pink shaded region). The intensity scale in the bottom row is normalized to the 80 K value. The dashed black lines in the bottom row indicate the angles parallel and perpendicular to the fastest growing lobe. Lattice point groups are labeled above.}
\end{figure*}

\section{Experimental details}
Single crystals of LiOsO$_3$ were prepared by solid state reaction under high pressure \cite{Shiferroelectriclikestructuraltransition2013a} and mechanically polished (roughness $<$0.5 $\mu$m) along their [121] surfaces. We chose this relatively low symmetry surface to enhance sensitivity to distortions both along and perpendicular to the polar $c$ axis [Fig.~\ref{fig1}(b)]. High surface qualities were verified by scanning electron microscopy and grain boundaries were carefully identified using electron backscatter diffraction [Fig.~\ref{fig1}(c)]. Light from a regeneratively amplified Ti:sapphire laser, which produces 80 fs pulses with central wavelength $\lambda$ = 800 nm at a repetition rate of 100 kHz, was focused at a variable angle of incidence ($\theta_i$) onto a 30 $\mu$m spot (FWHM) within a single [121] domain away from grain boundaries, with a fluence of 0.4 mJ/cm$^2$. Based on the reported optical constants of LiOsO$_3$ \cite{PadmanabhanLinearnonlinearoptical2018a}, the optical penetration depth at $\lambda$ = 800 nm is around 32 nm at normal incidence, and the propagation angle of the incident light inside the sample is around 0.8 $\theta_i$. The SHG-RA response was acquired by mechanically rotating the scattering plane about the sample surface normal and projecting the SHG light emitted at each scattering plane angle ($\varphi$) and emission angle ($\theta_e$) onto different positions on a two-dimensional CCD detector [Fig.~\ref{fig1}(d)] \cite{HarterHighspeedmeasurementrotational2015a,Torchinskylowtemperaturenonlinear2014b}. A typical raw SHG-RA data set from LiOsO$_3$ is shown in Fig.~\ref{fig1}(e), which manifests the symmetries of the [121] surface.

\section{Second harmonic generation results}
Below $T_c$ the electric-dipole (ED) contribution to SHG, governed by a third-rank susceptibility tensor $\chi_{ijk}^{ED}$ that relates the induced polarization at twice the frequency of the incident electric field via $P_{i}(2\omega)=\chi_{ijk}^{ED}E_{j}(\omega)E_{k}(\omega)$, becomes allowed due to the loss of inversion symmetry \cite{BoydNonlinearOptics2003,Perez-MatoCrystallographyOnlineBilbao2011}. Previous work has shown that $\chi_{ijk}^{ED}$ is linearly proportional to the polar order parameter and dominates the low-temperature SHG response from LiOsO$_3$ \cite{PadmanabhanLinearnonlinearoptical2018a}. However, in the centrosymmetric state above $T_c$, a finite SHG response can still arise from higher-multipole radiation, such as via an electric quadrupole (EQ) process $P_{i}(2\omega)=\chi_{ijkl}^{EQ}E_{j}(\omega)\nabla_{k}E_{l}(\omega)$, which is highly sensitive to symmetry-preserving lattice distortions \cite{RonDimensionalcrossoverlayered2019}, or a magnetic dipole process, which can be ruled out in our case \cite{EPAPS}. To detect and isolate the EQ from the ED contribution in LiOsO$_3$, we performed angle-of-incidence dependent SHG measurements in the S$\mathrm{_{in}}$-S$\mathrm{_{out}}$ geometry. In this geometry, the electric field polarizations of both the incident and detected SHG light are perpendicular to the scattering plane and maintain a fixed orientation relative to the sample as $\theta_{i}$ is varied. Therefore the ED contribution will be independent of $\theta_{i}$ whereas the EQ contribution will change with $\theta_{i}$ by virtue of its dependence on the light wave vector. Because our measurements are particularly sensitive to the Os-O sublattice symmetry at the chosen photon energies, the EQ contribution to the SHG intensity in the high-temperature centrosymmetric state should scale approximately as $\sin^{2}\theta_{i}$ \cite{EPAPS}. Therefore, we expect the EQ contribution to be significantly enhanced away from normal incidence \cite{ZhangDopingInducedSecondHarmonicGeneration2019a,BloembergenOpticalSecondHarmonicGeneration1968a}.

Figure~\ref{fig2}(a) shows that at $\theta_{i} = 0^{\circ}$, we detect no SHG signal above $T_c$ and a large intensity upturn just below $T_{c}$. The intensity scales with the square of the average Li atom displacement $\beta_{33}$ extracted from neutron diffraction data \cite{Shiferroelectriclikestructuraltransition2013a}, which is reproduced in Fig.~~\ref{fig2}(a). This is consistent with an ED dominated signal where all $\chi_{ijk}^{ED}$ elements scale linearly with the polar order parameter $\beta_{33}$, in accordance with a previous normal incidence SHG study \cite{PadmanabhanLinearnonlinearoptical2018a}. With a small increase of $\theta_{i}$ to 5$^{\circ}$, we detect a finite SHG intensity even at 300 K, which starts to grow below a characteristic temperature $T' \sim$ 230 K. The onset of the ED contribution at $T_c$ is slightly smeared out but remains identifiable via a kink in the SHG intensity [Fig.~\ref{fig2}(b)]. As $\theta_{i}$ increases further to 15$^{\circ}$, the SHG signal at 300 K becomes even stronger and again exhibits an upturn below $T'$. Now the kink at $T_c$ is completely obscured [Fig.~\ref{fig2}(c)]. No thermal hysteresis was observed in any measurement, consistent with a continuous phase transition.

The pronounced angle-of-incidence dependence of the high-temperature SHG intensity is incompatible with an ED process. This rules out the possibility of localized polar nano-regions, which generate SHG above $T_c$ in insulating ferroelectrics \cite{PugachevBrokenLocalSymmetry2012a}. Moreover, in the case of randomly oriented polar domains, one expects significant incoherent (hyper-Rayleigh) scattering \cite{Dolino, Melnikov}, manifested as the emission of second harmonic radiation over a broad range of angles around the specular direction $\theta_e = \theta_i$ [Fig.~\ref{fig1}(d)]. Figures~\ref{fig2}(d) and ~\ref{fig2}(e) show the $\theta_e$ dependence of the SHG intensity at a constant $\varphi$, which shows a sharp peak in the specular direction with no measurable broadening over a wide temperature range across $T_c$. We also found the ratio between the high-temperature SHG intensities at $\theta_{i} = 5^{\circ}$ and $\theta_{i} = 15^{\circ}$ [Figs~\ref{fig2}(b) and ~\ref{fig2}(c)] to be consistent with a $\sin^{2}\theta_{i}$ scaling (ignoring the weak $\theta_{i}$-dependence of the Fresnel coefficients over this range \cite{PadmanabhanLinearnonlinearoptical2018a}). Taken altogether, our data show that changes in the SHG signal above $T_c$ can be attributed to changes in the coherent EQ SHG response arising from lattice distortions.

\begin{figure}[t]
\includegraphics[width=1\columnwidth]{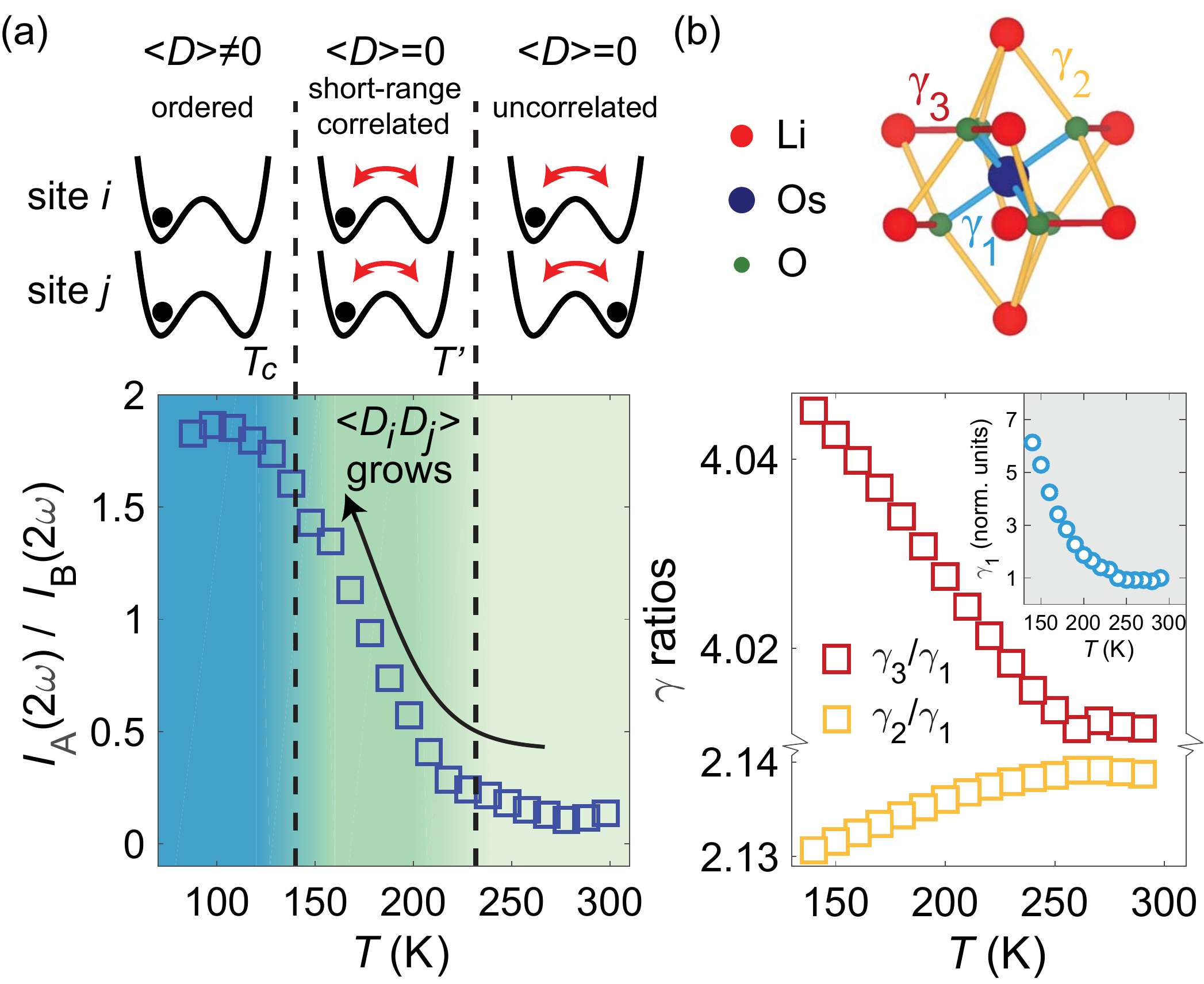}
\caption{\label{fig4} (a) The ratio of the SHG intensities at scattering plane angles A and B (see Fig.~\ref{fig3}) as a function of temperature. The evolution of dipolar correlations deduced from our data is illustrated above. Two neighboring sites are shown for each temperature range, and the black curves show the free energy potentials as a function of the Li polar displacement. For $T > T'$, local moments are already formed, but there is negligible correlation between sites. The local moment orientation, denoted by the location of the black dot in the double-well potential, fluctuates randomly. For $T_c < T < T'$, moments continue to fluctuate but short-range correlations develop. For $T < T_c$, moments become static long-range ordered. (b) The hyperpolarizabilities included in our SHBM are shown above. The extracted values of $\gamma_{2}$/$\gamma_{1}$, $\gamma_{3}$/$\gamma_{1}$, and $\gamma_{1}$ (inset) at each temperature obtained through fitting the SHG-RA patterns.}
\end{figure}

To understand the nature of the distortions above $T_c$, we measured the temperature dependence of the full SHG-RA patterns at $\theta_i$ = 15$^{\circ}$ where the EQ SHG contribution is pronounced. At $T$ = 290 K the SHG-RA pattern consists of four lobes with alternating amplitude. This fits the functional form predicted from a symmetry-based calculation of bulk EQ SHG from [121] oriented LiOsO$_3$ \cite{EPAPS}. No changes in either the magnitude or the shape of the SHG-RA pattern are observed down to $T'$ (Fig.~\ref{fig3}), consistent with a weak uniform thermal contraction of the lattice over this temperature range \cite{Shiferroelectriclikestructuraltransition2013a}. On the other hand, in the region bounded by $T'$ and $T_c$, we observe drastic changes to the shape of the patterns and the emergence of two additional intensity lobes. These changes do not lower the symmetry of the patterns and can be accounted for simply by adjusting the relative magnitudes of the $\chi_{ijkl}^{EQ}$ elements allowed in the nonpolar $\bar{3}m$ point group \cite{EPAPS}. We cannot reliably extract $\chi_{ijkl}^{EQ}$ values because the fits are under-determined, especially considering the $\chi_{ijkl}^{EQ}$'s can be complex for absorbing materials, but the important observation is their atypical nonuniform temperature dependence. This indicates that a nonuniform but symmetry-preserving lattice distortion, i.e., a polymorphous representation of the $R\bar{3}c$ structure, takes place between $T'$ and $T_c$, revealing an extended critical region over which $\langle D_{i}D_{j}\rangle$ grows from dissimilar Li displacements that are correlated over the probe volume. Upon further cooling below $T_c$, the SHG intensity continues to rise, but the shape of the SHG-RA patterns does not evolve appreciably. This is to be expected once ED SHG starts to dominate the signal because all elements of $\chi_{ijk}^{ED}$ are linearly proportional to the polar order parameter. This means that the relative magnitudes of all $\chi_{ijk}^{ED}$ elements are constant, which fixes the pattern shape. The full temperature evolution of the pattern shape can be quantified by plotting the intensity ratio of lobes A and B defined in Fig.~\ref{fig3}. As shown in Fig.~\ref{fig4}(a), this intensity ratio grows steeply between $T'$ and $T_c$ but is temperature independent outside this window, in stark contrast to the behaviors of the intensities alone (Fig.~\ref{fig2}).

\section{Bond model analysis}
We explored the microscopic origin of the SHG-RA pattern evolution by considering a simplified hyper-polarizable bond model (SHBM) \cite{PowellSimplifiedbondhyperpolarizabilitymodel2002a,BauerBulkquadrupolecontribution2017a}. This model treats the crystal as an array of charged anharmonic oscillators centered at the chemical bonds, with motion constrained along the bond directions. The nonlinear susceptibilities of individual oscillators are calculated by solving classical equations of motion, and then summed together to form a bulk nonlinear susceptibility. An expression for $\chi_{ijkl}^{EQ}$ derived from a SHBM has been shown to take the form $\chi_{ijkl}^{EQ}\propto\sum\limits_{n}\gamma^{\omega}_{n}\gamma^{2\omega}_{n}(\hat{b}_{n}\otimes\hat{b}_{n}\otimes\hat{b}_{n}\otimes\hat{b}_{n})_{ijkl}$, where $\gamma^{\omega}_{n}$ and $\gamma^{2\omega}_{n}$ are the linear and second-order (hyper-) polarizabilities of the $n^{\mathrm{th}}$ bond, $\hat{b}_n$ is a unit vector pointing along the $n^{\mathrm{th}}$ bond, and all bond charges are assumed equal. For LiOsO$_3$, we included all three types of bonds in the unit cell: the Os-O bonds, the long Li-O bonds between adjacent $c$ planes, and the short Li-O bonds in the same $c$ planes \cite{BenedekFerroelectricmetalsreexamined2016}, with hyper-polarizabilities $\gamma_{1}$, $\gamma_{2}$, and $\gamma_{3}$, respectively [Fig.~\ref{fig4}(b)]. No measurable change in the linear optical response at 1.5 eV or 3 eV has been detected over the temperature range studied here \cite{LoVecchioElectroniccorrelationsferroelectric2016,LauritaEvidenceweaklycoupled2019}. The change in bond directions $\hat{b}_n$ alone based on neutron diffraction data \cite{Shiferroelectriclikestructuraltransition2013a} is also too small to account for the observed changes in $\chi_{ijkl}^{EQ}$ within the SHBM \cite{EPAPS}. Therefore, the $\gamma^{\omega}_{n}$ and $\hat{b}_n$ are held fixed in our analysis, and the hyper-polarizabilities are the only free parameters. Best fits of this SHBM to our SHG-RA data are shown in the bottom panels of Fig.~\ref{fig3}, from which we obtain the temperature dependence of $\gamma_{1}$, $\gamma_{2}$, and $\gamma_{3}$. We find that while all hyper-polarizability values grow upon cooling, a small decrease (increase) in $\gamma_{2}$/$\gamma_{1}$ ($\gamma_{3}$/$\gamma_{1}$) occurring over the critical region is responsible for the large shape changes observed in the SHG-RA data [Fig.~\ref{fig4}(b)]. Knowing that uniform thermal contraction above $T'$ does not affect the shape of the SHG-RA patterns, we can assume that the scaling between the change in bond hyper-polarizability and the change in bond length is the same for different bonds. Therefore, the critical region must be characterized by nonuniform changes in the short and long Li-O bond lengths.

\section{Discussion and conclusions}
A microscopic picture of the temperature evolution of dipole correlations in LiOsO$_3$ deduced from our electrostriction measurements is presented in the top panels of Fig.~\ref{fig4}(a). Our finding of an extended critical region across $T_c < T < T'$ in LiOsO$_3$ corroborates the predominantly order-disorder character of the polar transition and points to the presence of significant dipolar correlations developing below $T'$. This may explain the low entropy loss across $T_c$ found in specific heat \cite{Shiferroelectriclikestructuraltransition2013a}, the unusual increase in linewidth of the $^{3}E_g$ Li in-plane vibrational mode below $\sim T'$ observed by Raman scattering \cite{JinRamaninterrogationferroelectric2019}, as well as the weakly nonuniform changes occurring in the $a$ and $c$ lattice parameters below $\sim T'$ measured by neutron diffraction \cite{Shiferroelectriclikestructuraltransition2013a}. The fact that previous Curie-Weiss analyses of dielectric susceptibility data were performed inside the critical region may also be responsible for the unusually low value of the extracted Curie-Weiss temperature \cite{JinRamaninterrogationferroelectric2019}. An extended critical region is not accounted for in existing 3D Ising-type models with purely ferroic interactions, which were proposed based on density functional theory calculations \cite{LiuMetallicferroelectricityinduced2015a}. There is also no evidence for spatially anisotropic interactions in the Hamiltonian that would lead to effectively lower dimensionality and hence enhanced fluctuations \cite{LiuMetallicferroelectricityinduced2015a}. Instead, it is possible that enhanced fluctuations arise from the geometric origin of Li off-centering --- competing interactions from different signs of $J_{ij}$ arising owing to different neighbor distances. The nonuniform change in the short and long Li-O bonds serves to partially relieve the multiple nearly degenerate displacements. Our findings should guide further theoretical work in search of a detailed microscopic mechanism, which can in turn lead to more refined and quantitative models of the SHG response that go beyond our phenomenological treatment. More generally, our technique can be applied broadly to understand critical phenomena in geometric ferroelectrics, for which many polar metals are an effective subclass, as well as other metals with broken inversion symmetry, such as topological semimetals. It would also be interesting to explore signatures of critical fluctuations in other related nonlinear optical responses such as the photogalvanic effect, which has recently been shown to be a symmetry-sensitive probe of noncentrosymmetric topological semimetals \cite{GedikTaAs,OrensteinRhSi}.\\\\

\begin{acknowledgments}
We thank A. Boothroyd and D. Khalyavin for sharing their neutron diffraction data. We also acknowledge useful discussions with S. Biswas, J. S. Lee, P. A. Lee, C. Li, C. Ma, A. Ron and J. Schmehr. This work was supported by the U.S. Department of Energy under Grant No. DE SC0010533. D.H. also acknowledges funding from the David and Lucile Packard Foundation and support for instrumentation from the Institute for Quantum Information and Matter, an NSF Physics Frontiers Center (PHY-1733907). N.J.L. acknowledges partial support from the IQIM postdoctoral fellowship. D.P. and J.M.R. were supported by ARO (Award No. W911NF-15-1-0017). Y.G.S. was supported by the National Key Research and Development Program of China (No. 2017YFA0302901 and 2016YFA0300604), as well as by the K. C. Wong Education Foundation (GJTD-2018-01).
\end{acknowledgments}



%

\end{document}



\title{SUPPLEMENTAL MATERIAL \\ \vspace{5mm} Evidence for an extended critical fluctuation region above the polar ordering transition in LiOsO$_{3}$}

\affiliation{Department of Physics, California Institute of Technology, Pasadena, California 91125, USA\looseness=-1}
\affiliation{Institute for Quantum Information and Matter, California Institute of Technology, Pasadena, California 91125, USA\looseness=-1}
\affiliation{Department of Physics, University of Michigan, Ann Arbor, Michigan 48109, USA\looseness=-1}
\affiliation{London Centre for Nanotechnology and Department of Physics and Astronomy, University College London, London WC1E 6BT, UK\looseness=-1}
\affiliation{Department of Materials Science and Engineering, Northwestern University, Illinois 60208-3108, USA\looseness=-1}
\affiliation{Research Center for Functional Materials, National Institute for Materials Science, 1-1 Namiki, Tsukuba, Ibaraki 305-0044, Japan\looseness=-1}
\affiliation{Institute of Physics, Chinese Academy of Sciences, Beijing 100190, China\looseness=-1}

\author{Jun-Yi Shan} \affiliation{Department of Physics, California Institute of Technology, Pasadena, California 91125, USA\looseness=-1}\affiliation{Institute for Quantum Information and Matter, California Institute of Technology, Pasadena, California 91125, USA\looseness=-1}
\author{A. \surname{de la Torre}}\affiliation{Department of Physics, California Institute of Technology, Pasadena, California 91125, USA\looseness=-1}\affiliation{Institute for Quantum Information and Matter, California Institute of Technology, Pasadena, California 91125, USA\looseness=-1}
\author{N. J. Laurita}\affiliation{Department of Physics, California Institute of Technology, Pasadena, California 91125, USA\looseness=-1}\affiliation{Institute for Quantum Information and Matter, California Institute of Technology, Pasadena, California 91125, USA\looseness=-1}
\author{L. Zhao}  \affiliation{Department of Physics, University of Michigan, Ann Arbor, Michigan 48109, USA\looseness=-1}
\author{C. D. Dashwood} \affiliation{London Centre for Nanotechnology and Department of Physics and Astronomy, University College London, London WC1E 6BT, UK\looseness=-1}
\author{D. Puggioni} \affiliation{Department of Materials Science and Engineering, Northwestern University, Illinois 60208-3108, USA\looseness=-1}
\author{C. X. Wang} \affiliation{Institute of Physics, Chinese Academy of Sciences, Beijing 100190, China\looseness=-1}
\author{K. Yamaura} \affiliation{Research Center for Functional Materials, National Institute for Materials Science, 1-1 Namiki, Tsukuba, Ibaraki 305-0044, Japan\looseness=-1}
\author{Y. Shi} \affiliation{Institute of Physics, Chinese Academy of Sciences, Beijing 100190, China\looseness=-1}
\author{J. M. Rondinelli} \affiliation{Department of Materials Science and Engineering, Northwestern University, Illinois 60208-3108, USA\looseness=-1}
\author{D. Hsieh}
\affiliation{Department of Physics, California Institute of Technology, Pasadena, California 91125, USA\looseness=-1}\affiliation{Institute for Quantum Information and Matter, California Institute of Technology, Pasadena, California 91125, USA\looseness=-1}

\date{\today}

\pacs{Valid PACS appear here}
\maketitle

\renewcommand{\thesection}{S\arabic{section}}
\renewcommand{\thetable}{S\arabic{table}}
\renewcommand{\thefigure}{S\arabic{figure}}
\renewcommand*{\citenumfont}[1]{S#1}
\renewcommand*{\bibnumfmt}[1]{[S#1]}

\section{Determination of $T'$}

\begin{figure}[h]
\includegraphics[width=1\columnwidth]{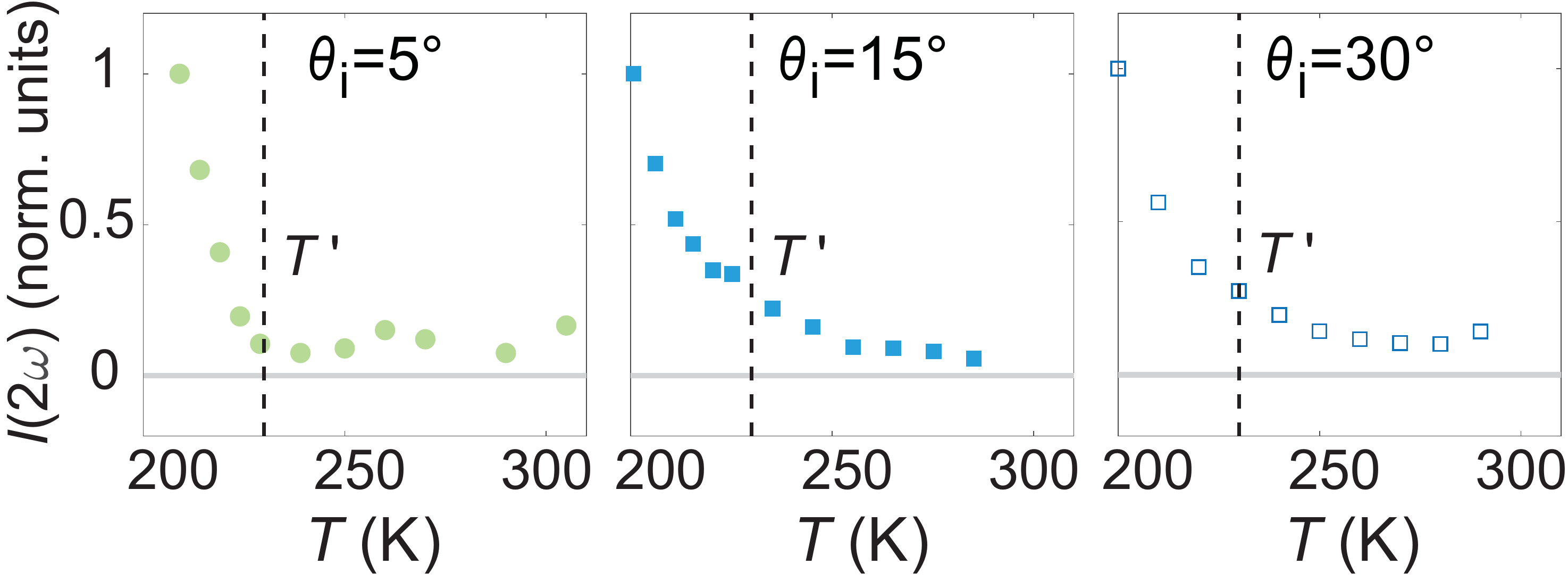}
\caption{\label{tprime} Zoom-in on the SHG intensity near $T'$ (vertical dashed line) for $\theta_{i}$ = 5$^{\circ}$, 15$^{\circ}$, and 30$^{\circ}$. All data are normalized to their values at $T$ = 200 K.}
\end{figure}

We defined the value of $T'$ as the characteristic temperature below which the EQ contribution to the SHG intensity starts to grow. It is not sharply defined since no phase transition occurs at $T'$. As shown in Fig.~\ref{tprime}, we consistently observe that $T' \sim$ 230 K independent of the angle of incidence $\theta_i$. Note that $T'$ is not detectable at normal incidence because the EQ contribution vanishes.

\section{SHG-RA data for different polarization geometries}

\begin{figure*}[t]
\includegraphics[width=1\columnwidth]{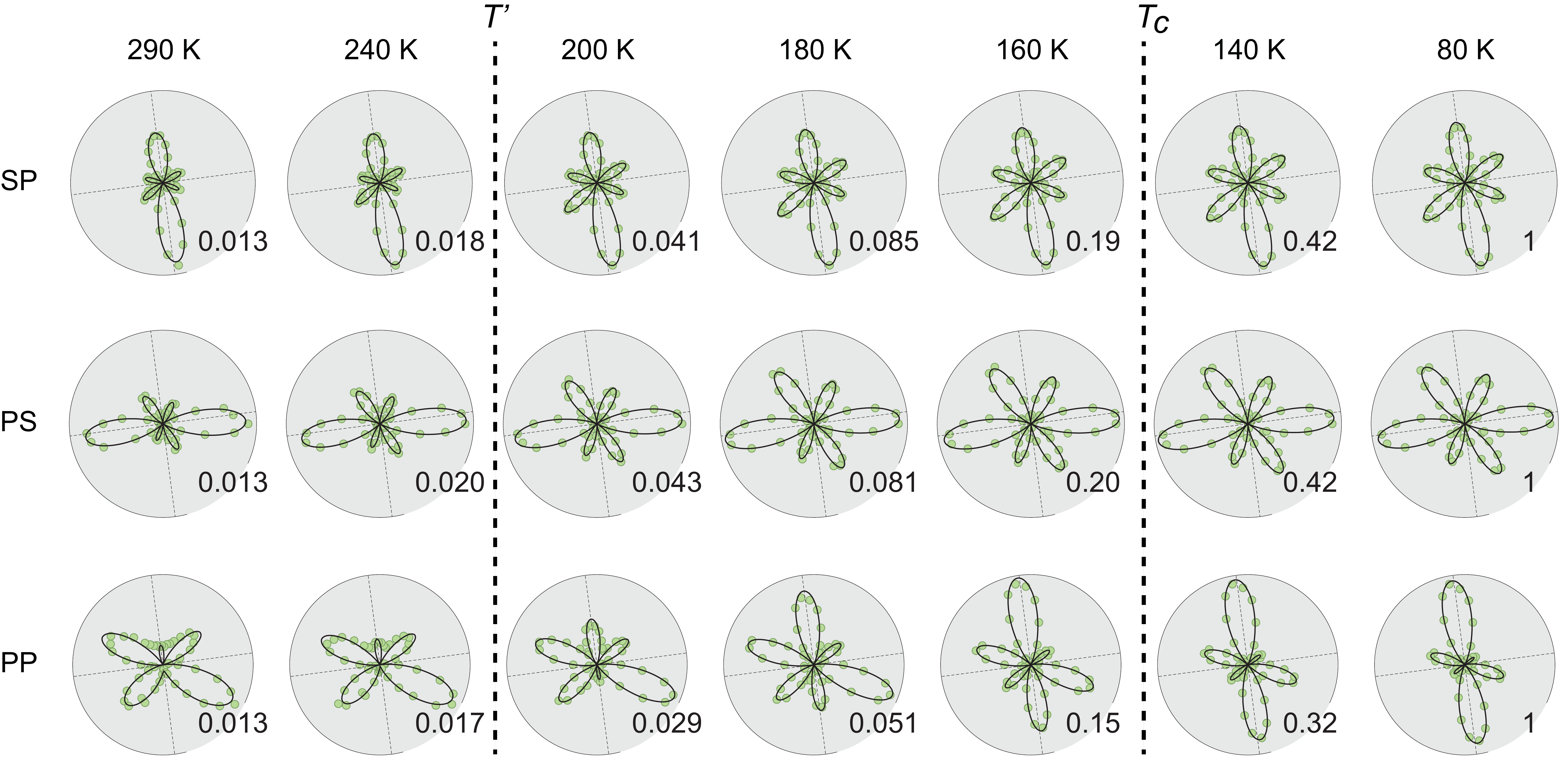}
\caption{\label{other} Polar plots of the SHG-RA data (green circles) from LiOsO$_{3}$ [121] at selected temperatures taken under S$\mathrm{_{in}}$-P$\mathrm{_{out}}$, P$\mathrm{_{in}}$-S$\mathrm{_{out}}$, and P$\mathrm{_{in}}$-P$\mathrm{_{out}}$ polarization geometries. Fits to an EQ SHG response for an $R\bar{3}c$ space group ($\bar{3}m$ point group) as described in Section S4 are superposed as black lines. For the 80 K fit we included an additional ED SHG contribution for an $R3c$ space group ($3m$ point group). The intensity scale is normalized to the 80 K value.}
\end{figure*}

The SHG-RA data shown in the main text are taken under S$\mathrm{_{in}}$-S$\mathrm{_{out}}$ polarization geometry. Figure~\ref{other} shows the SHG-RA data sets taken under the alternate S$\mathrm{_{in}}$-P$\mathrm{_{out}}$, P$\mathrm{_{in}}$-S$\mathrm{_{out}}$, and P$\mathrm{_{in}}$-P$\mathrm{_{out}}$ geometries. Consistent with the S$\mathrm{_{in}}$-S$\mathrm{_{out}}$ data reported in the main text, they all exhibit clear shape changes over the temperature interval $\it{T'}>\it{T}>\it{T_{c}}$.

\section{Fits to MD SHG processes}

The pronounced angle-of-incidence dependence of the SHG response in the high temperature centrosymmetric state (main text Fig. 2) is incompatible with an ED process, but may arise from bulk magnetic-dipole (MD) or bulk EQ processes. In addition to the EQ response discussed in the main text, we considered the two possible MD SHG contributions of the type $\it{M_{i}}(2\omega)=\chi_{\it{ijk}}^{\it{MD,1}}\it{E_{j}}(\omega)\it{E_{k}}(\omega)$ and $\it{P_{i}}(2\omega)=\chi_{\it{ijk}}^{\it{MD,2}}\it{E_{j}}(\omega)\it{H_{k}}(\omega)$. In the $R\bar{3}$c space group, there are two nonzero independent tensor elements for $\chi_{\it{ijk}}^{\it{MD,1}}$, which are \cite{Perez-MatoCrystallographyOnlineBilbao2011} $-\chi^{MD,1}_{xxx}= \chi^{MD,1}_{xyy}= \chi^{MD,1}_{yxy}= \chi^{MD,1}_{yyx}$,  and $-\chi^{MD,1}_{xyz}= -\chi^{MD,1}_{xzy}= \chi^{MD,1}_{yxz}= \chi^{MD,1}_{yzx}$. The $z$ direction is along the crystallographic $c$ axis, and the $x$ direction is perpendicular to the glide mirror plane. For LiOsO$_3$ [121], the dependence of this MD SHG intensity in S$\mathrm{_{in}}$-S$\mathrm{_{out}}$ geometry on the angle of incidence $\theta_i$ and scattering plane angle $\varphi$ is given by $I^{MD,1}(2\omega)=[\sin{\theta_{i}}(a_{1}\sin^{2}\varphi+a_{2}\sin\varphi\cos\varphi+a_{3}\cos^{2}\varphi)+\cos{\theta_{i}}(a_{4}\sin^{3}\varphi+a_{5}\sin^{2}\varphi\cos\varphi+a_{6}\sin\varphi\cos^{2}\varphi+a_{7}\cos^{3}\varphi)]^{2}$, where
\begin{align*}
a_{1}&=-0.27\chi_{xxx}^{MD,1} - 0.22\chi_{xyz}^{MD,1},\\
a_{2}&=0.82\chi_{xxx}^{MD,1} - 0.38\chi_{xyz}^{MD,1},\\
a_{3}&=0.12\chi_{xxx}^{MD,1} + 0.22\chi_{xyz}^{MD,1},\\
a_{4}&=0.05\chi_{xxx}^{MD,1} + 0.38\chi_{xyz}^{MD,1},\\
a_{5}&=-2.39\chi_{xxx}^{MD,1} + 0.82\chi_{xyz}^{MD,1},\\
a_{6}&=0.02\chi_{xxx}^{MD,1} + 0.38\chi_{xyz}^{MD,1},\\
a_{7}&=0.74\chi_{xxx}^{MD,1} + 0.82\chi_{xyz}^{MD,1}.\\
\end{align*}

\begin{figure}[h]
\includegraphics[width=1\columnwidth]{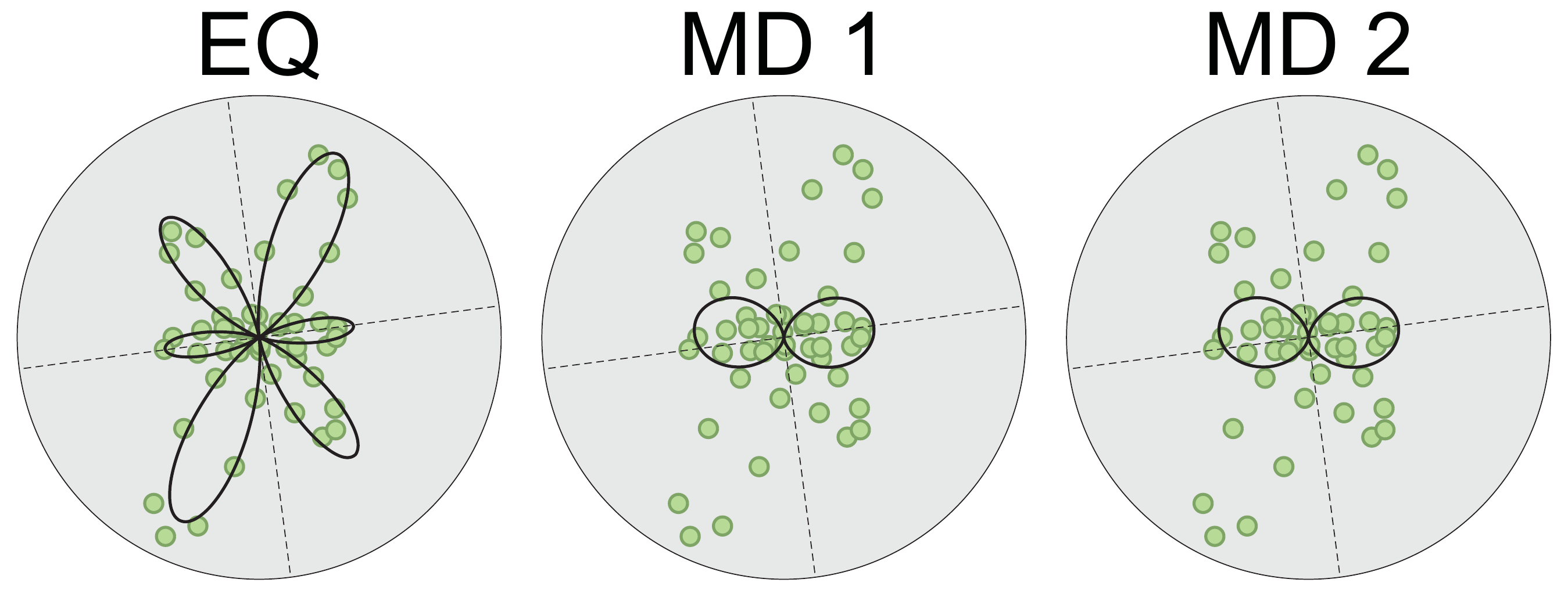}
\caption{\label{md} SHG-RA data taken in S$\mathrm{_{in}}$-S$\mathrm{_{out}}$ geometry (green circles) from LiOsO$_{3}$ [121] at 200 K and the best fits to the EQ, MD 1, and MD 2 SHG processes (black lines).}
\end{figure}

There are four nonzero independent tensor elements for $\chi_{\it{ijk}}^{\it{MD,2}}$, which are $-\chi^{MD,2}_{xxx}=\chi^{MD,2}_{xyy}=\chi^{MD,2}_{yxy}=\chi^{MD,2}_{yyx}$, $-\chi^{MD,2}_{xyz}=\chi^{MD,2}_{yxz}$, $-\chi^{MD,2}_{xzy}=\chi^{MD,2}_{yzx}$, and $-\chi^{MD,2}_{zxy}=\chi^{MD,2}_{zyx}$.  The expression for this MD SHG intensity in S$\mathrm{_{in}}$-S$\mathrm{_{out}}$ geometry is given by $I^{MD,2}(2\omega)=[\sin{\theta_{i}}(a_{8}\sin^{2}\varphi+a_{9}\sin\varphi\cos\varphi+a_{10}\cos^{2}\varphi)+\cos{\theta_{i}}(a_{11}\sin^{3}\varphi+a_{12}\sin^{2}\varphi\cos\varphi+a_{13}\sin\varphi\cos^{2}\varphi+a_{14}\cos^{3}\varphi)]^{2}$, where
\begin{align*}
a_{8}&=-0.27\chi_{xxx}^{MD,2} + 0.11\chi_{xzy}^{MD,2} + 0.11\chi_{zxy}^{MD,2},\\
a_{9}&=0.83\chi_{xxx}^{MD,2}  - 0.10\chi_{xzy}^{MD,2} + 0.48\chi_{zxy}^{MD,2},\\
a_{10}&=0.12\chi_{xxx}^{MD,2} - 0.11\chi_{xzy}^{MD,2} - 0.11\chi_{zxy}^{MD,2},\\
a_{11}&=0.05\chi_{xxx}^{MD,2} - 0.19\chi_{xzy}^{MD,2} - 0.19\chi_{zxy}^{MD,2},\\
a_{12}&=-2.39\chi_{xxx}^{MD,2}  - 0.41\chi_{xzy}^{MD,2} - 0.41\chi_{zxy}^{MD,2},\\
a_{13}&=0.02\chi_{xxx}^{MD,2} - 0.19\chi_{zxy}^{MD,2}  - 0.19\chi_{xzy}^{MD,2},\\
a_{14}&=0.74\chi_{xxx}^{MD,2} - 0.41\chi_{xzy}^{MD,2} - 0.41\chi_{zxy}^{MD,2}.\\
\end{align*}

As shown in Fig. \ref{md}, neither of these two MD processes fits our high-temperature SHG-RA data, due to insufficient number of independent fitting parameters. Thus, we conclude that the EQ SHG contribution dominates the high-temperature SHG signal.

\section{EQ and ED SHG susceptibility tensors}

We considered an EQ SHG process of the form $P_{i}=\chi_{ijkl}^{EQ}E_{j}\nabla_{k}E_{l}$. In the $R\bar{3}$c space group, the $\chi_{ijkl}^{EQ}$ susceptibility tensor has 11 nonzero independent tensor elements: \cite{BoydNonlinearOptics2003} $\chi^{EQ}_{xxxx}=\chi^{EQ}_{yyyy}$, $\chi^{EQ}_{xxyy}=\chi^{EQ}_{yyxx}$, $\chi^{EQ}_{xxzz}=\chi^{EQ}_{yyzz}=\chi^{EQ}_{xzzx}=\chi^{EQ}_{yzzy}$, $\chi^{EQ}_{xxyz}=\chi^{EQ}_{yxxz}=\chi^{EQ}_{xyxz}=-\chi^{EQ}_{yyyz}=-\chi^{EQ}_{yzyy}=\chi^{EQ}_{xzxy}=\chi^{EQ}_{xzyx}=\chi^{EQ}_{yzxx}$, $\chi^{EQ}_{xxzy}=\chi^{EQ}_{yxzx}=\chi^{EQ}_{xyzx}=-\chi^{EQ}_{yyzy}$, $\chi^{EQ}_{xyxy}=\chi^{EQ}_{yxyx}$, $\chi^{EQ}_{xzxz}=\chi^{EQ}_{yzyz}$, $\chi^{EQ}_{zxxy}=\chi^{EQ}_{zxyx}=\chi^{EQ}_{zyxx}=-\chi^{EQ}_{zyyy}$, $\chi^{EQ}_{zxzx}=\chi^{EQ}_{zyzy}$, $\chi^{EQ}_{zxxz}=\chi^{EQ}_{zyyz}=\chi^{EQ}_{zzxx}=\chi^{EQ}_{zzyy}$, and $\chi^{EQ}_{zzzz}$. Also $\chi^{EQ}_{xyyx}=\chi^{EQ}_{yxxy}=\chi^{EQ}_{xxxx}-\chi^{EQ}_{xxyy}-\chi^{EQ}_{xyxy}$. For LiOsO$_3$ [121], the dependence of the SHG intensity in S$\mathrm{_{in}}$-S$\mathrm{_{out}}$ geometry on the angle of incidence $\theta_i$ and scattering plane angle $\varphi$ is given by $I^{EQ}(2\omega)=[\sin{\theta_{i}}(c_{1}\sin^{4}\varphi+c_{2}\sin^{3}\varphi\cos\varphi+c_{3}\sin^{2}\varphi\cos^{2}\varphi+c_{4}\sin\varphi\cos^{3}\varphi+c_{5}\cos^{4}\varphi)+\cos{\theta_{i}}(c_{6}\sin^{3}\varphi+c_{7}\sin^{2}\varphi\cos\varphi+c_{8}\sin\varphi\cos^{2}\varphi+c_{9}\cos^{3}\varphi)]^{2}$, where\\\\
$c_{1}=0.10\chi_{xxxx}^{EQ} - 0.21\chi_{xxzz}^{EQ} + 0.01\chi_{xzxz}^{EQ} - 0.37\chi_{yyyz}^{EQ} + 0.09\chi_{yyzy}^{EQ} - 0.10\chi_{zxzx}^{EQ} - 0.19\chi_{zyyy}^{EQ} + 0.01\chi_{zzxx}^{EQ} - 0.01\chi_{zzzz}^{EQ}$,\\\\
$c_{2}=0.15\chi_{xxxx}^{EQ} - 0.31\chi_{xxzz}^{EQ} + 0.03\chi_{xzxz}^{EQ} - 0.85\chi_{yyyz}^{EQ} - 1.25\chi_{yyzy}^{EQ} - 0.15\chi_{zxzx}^{EQ} - 0.42\chi_{zyyy}^{EQ} + 0.07\chi_{zzxx}^{EQ} - 0.03\chi_{zzzz}^{EQ}$,\\\\
$c_{3}=-0.06\chi_{xxxx}^{EQ} + 0.12\chi_{xxzz}^{EQ} + 0.06\chi_{xzxz}^{EQ} + 0.96\chi_{yyyz}^{EQ} + 0.63\chi_{yyzy}^{EQ} +  0.06 \chi_{zxzx}^{EQ} + 0.48\chi_{zyyy}^{EQ} + 0.12\chi_{zzxx}^{EQ} - 0.06\chi_{zzzz}^{EQ}$,\\\\
$c_{4}=0.17\chi_{xxxx}^{EQ}-  0.34\chi_{xxzz}^{EQ} + 0.02\chi_{xzxz}^{EQ} + 2.23\chi_{yyyz}^{EQ} + 0.29\chi_{yyzy}^{EQ} - 0.17\chi_{zxzx}^{EQ} + 1.12\chi_{zyyy}^{EQ} + 0.04\chi_{zzxx}^{EQ} - 0.02\chi_{zzzz}^{EQ}$,\\\\
$c_{5}=-0.08\chi_{xxxx}^{EQ} + 0.17\chi_{xxzz}^{EQ} - 0.03\chi_{xzxz}^{EQ} - 0.05\chi_{yyyz}^{EQ} - 0.15\chi_{yyzy}^{EQ} + 0.08\chi_{zxzx}^{EQ} - 0.02\chi_{zyyy}^{EQ} - 0.05\chi_{zzxx}^{EQ} + 0.03\chi_{zzzz}^{EQ}$,\\\\
$c_{6}=-0.18\chi_{xxxx}^{EQ} + 0.36\chi_{xxzz}^{EQ} - 0.01\chi_{xzxz}^{EQ} - 0.19\chi_{yyyz}^{EQ} - 0.15\chi_{yyzy}^{EQ} + 0.18\chi_{zxzx}^{EQ} - 0.10\chi_{zyyy}^{EQ} - 0.02\chi_{zzxx}^{EQ} + 0.01\chi_{zzzz}^{EQ}$,\\\\
$c_{7}=-0.35\chi_{xxxx}^{EQ} + 0.70\chi_{xxzz}^{EQ} - 0.06\chi_{xzxz}^{EQ} - 0.60\chi_{yyyz}^{EQ} + 2.09\chi_{yyzy}^{EQ} + 0.35\chi_{zxzx}^{EQ} - 0.30\chi_{zyyy}^{EQ} - 0.13\chi_{zzxx}^{EQ} + 0.06\chi_{zzzz}^{EQ}$,\\\\
$c_{8}=-0.05\chi_{xxxx}^{EQ} + 0.11\chi_{xxzz}^{EQ} - 0.14\chi_{xzxz}^{EQ} - 0.20\chi_{yyyz}^{EQ} - 0.12\chi_{yyzy}^{EQ} + 0.05\chi_{zxzx}^{EQ} - 0.10\chi_{zyyy}^{EQ} - 0.27\chi_{zzxx}^{EQ} + 0.14\chi_{zzzz}^{EQ}$,\\\\
$c_{9}=-0.31\chi_{xxxx}^{EQ} + 0.63\chi_{xxzz}^{EQ} - 0.10\chi_{xzxz}^{EQ} + 0.38\chi_{yyyz}^{EQ} - 0.55\chi_{yyzy}^{EQ} + 0.31\chi_{zxzx}^{EQ} + 0.20\chi_{zyyy}^{EQ} - 0.20\chi_{zzxx}^{EQ} + 0.10\chi_{zzzz}^{EQ}$.\\

We can fit all SHG-RA data sets in the region $\it{T'}>\it{T}>\it{T_{c}}$ shown in Fig. 3 of the main text simply by adjusting the magnitudes of the allowed tensor elements.

We note that even though the EQ SHG response has both a $\cos{\theta_{i}}$ and $\sin{\theta_{i}}$ dependence, experimentally we find that the SHG intensity in the centrosymmetric phase becomes vanishingly small at normal incidence. Therefore the $\sin{\theta_{i}}$ term dominates. This is expected for the following reason. The [121] oriented LiOsO$_3$ crystal in the centrosymmetric phase does not have twofold rotational ($C_2$) symmetry about the surface normal due to the positioning of the Li atoms [see main text Fig. 1(b)]. However the Os-O sublattice does have $C_2$ symmetry. Our SHG measurement uses an incident photon energy of 1.5 eV and an SHG photon energy of 3 eV, so the intermediate and final states of the SHG process are dominated by Os-5$d$ and O-2$p$ states \cite{LiuMetallicferroelectricityinduced2015a}. Therefore we should be primarily sensitive to the Os-O sublattice. This is consistent with the near $C_2$ symmetry we observe in our SHG-RA data. Since the coefficients of the $\cos{\theta_{i}}$ term all break $C_2$, they must be significantly smaller than the coefficients of the $\sin{\theta_{i}}$ term, which preserves $C_2$, at the photon energies we use.

Below $\it{T_{c}}$ an additional bulk ED SHG contribution is allowed, which interferes with the EQ contribution. In the $R3c$ space group, the $\chi_{ijk}^{ED}$ susceptibility tensor has four nonzero independent tensor elements \cite{BoydNonlinearOptics2003}, $\chi^{ED}_{xxy}=\chi^{ED}_{xyx}=\chi^{ED}_{yxx}=-\chi^{ED}_{yyy}$, $\chi^{ED}_{xxz}=\chi^{ED}_{xzx}=\chi^{ED}_{yyz}=\chi^{ED}_{yzy}$, $\chi^{ED}_{zxx}=\chi^{ED}_{zyy}$, and $\chi^{ED}_{zzz}$. For LiOsO$_3$ [121], the expression for the ED SHG intensity in S$\mathrm{_{in}}$-S$\mathrm{_{out}}$ geometry is given by $I^{ED}(2\omega)=(d_{1}\sin^{3}\varphi+d_{2}\sin^{2}\varphi\cos\varphi+d_{3}\sin\varphi\cos^{2}\varphi+d_{4}\cos^{3}\varphi)^{2}$, where
\begin{align*}
d_{1}&=0.43\chi_{xxz}^{ED} - 0.18\chi_{yyy}^{ED} + 0.21\chi_{zxx}^{ED} + 0.01\chi_{zzz}^{ED},\\
d_{2}&=0.93\chi_{xxz}^{ED} + 2.48\chi_{yyy}^{ED} + 0.41\chi_{zxx}^{ED} + 0.07\chi_{zzz}^{ED},\\
d_{3}&=0.13\chi_{xxz}^{ED} - 0.15\chi_{yyy}^{ED} + 0.06\chi_{zxx}^{ED} + 0.16\chi_{zzz}^{ED},\\
d_{4}&=0.74\chi_{xxz}^{ED} - 0.65\chi_{yyy}^{ED} + 0.37\chi_{zxx}^{ED} + 0.12\chi_{zzz}^{ED}.\\
\end{align*}

\begin{figure*}[h]
\includegraphics[width=1\columnwidth]{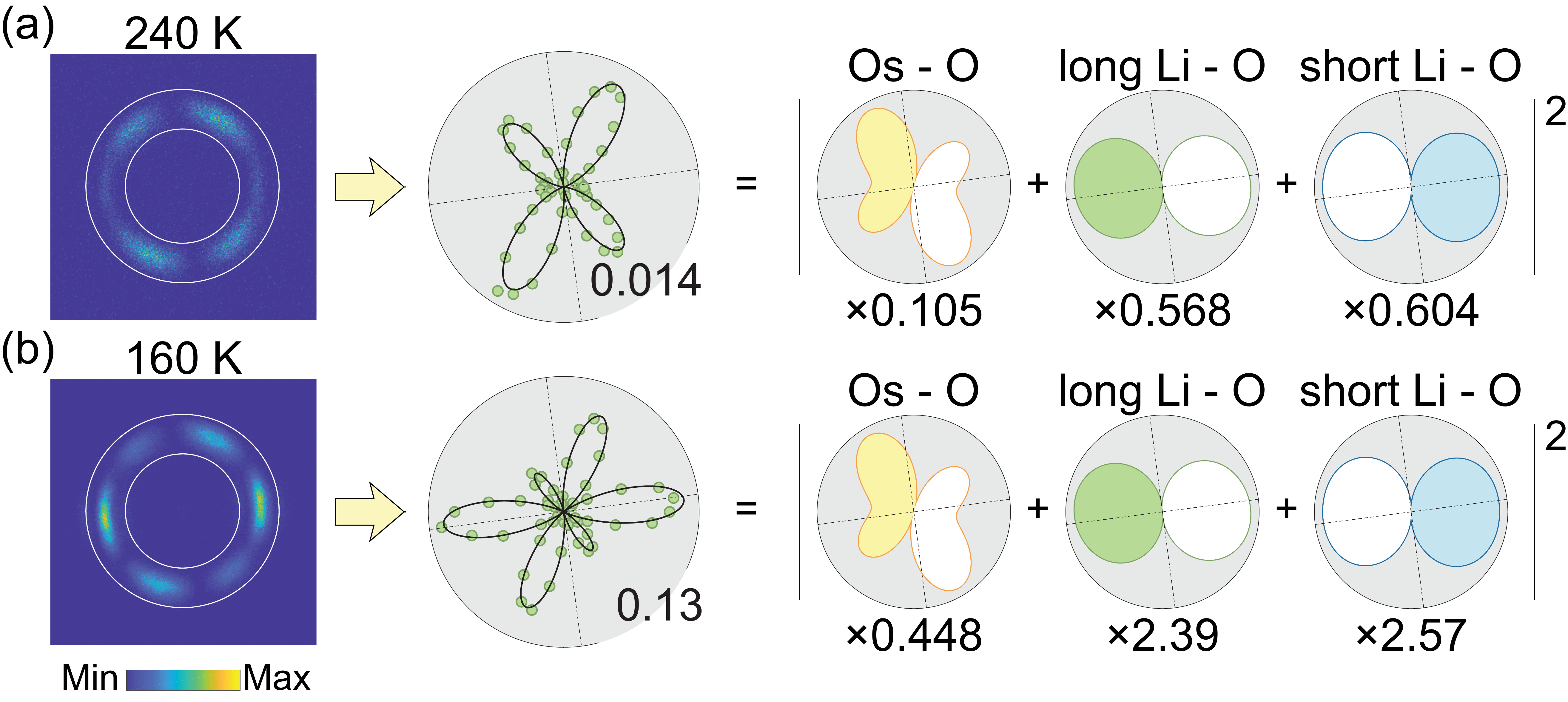}
\caption{\label{fig4_1} Example fits of the SHG-RA data at (a) 240 K and (b) 160 K to the SHBM. The first column shows the raw data. The second column shows the angular dependence of the SHG intensity obtained by radially integrating the raw data between the region bounded by the concentric white circles. Black curves are the SHBM fits. The third column shows the contributions from the Os-O bonds, the long Li-O bonds, and the short Li-O bonds obtained from fitting, with their fitted magnitudes written below. Shaded (unshaded) areas indicate positive (negative) amplitudes.}
\end{figure*}
\section{Details of the simplified hyper-polarizable bond model}
In the SHBM, the EQ susceptibility tensor can be expressed as $\chi_{ijkl}^{EQ}\propto\sum\limits_{n}\gamma_{n}^{\omega}\gamma_{n}^{2\omega}(\hat{\textbf{b}}_{n}\otimes\hat{\textbf{b}}_{n}\otimes\hat{\textbf{b}}_{n}\otimes\hat{\textbf{b}}_{n})_{ijkl}$, where $\gamma_{n}^{\omega}$ and $\gamma_{n}^{2\omega}$ are the linear and second-order (hyper-) polarizabilities of the $n^{\mathrm{th}}$ bond and $\hat{\textbf{b}}_{n}$ is a unit vector along the direction of the $n^{\mathrm{th}}$ bond \cite{PowellSimplifiedbondhyperpolarizabilitymodel2002a,BauerBulkquadrupolecontribution2017a}. To fit our SHG-RA data we include all three types of bonds in the unit cell of LiOsO$_{3}$: the Os-O bonds, the long Li-O bonds between adjacent $c$ planes, and the short Li-O bonds in the same $c$ planes \cite{BenedekFerroelectricmetalsreexamined2016}. Based on the lack of marked temperature dependence in the reported linear optical response \cite{LoVecchioElectroniccorrelationsferroelectric2016,LauritaEvidenceweaklycoupled2019}, we assume that the values of $\gamma_{n}^{\omega}$ are temperature independent. We also keep $\hat{\textbf{b}}_{n}$ values constant, leaving only the $\gamma_{n}^{2\omega}$ as fit parameters (Fig.~\ref{fig4_1}). The $\hat{\textbf{b}}_{n}$'s of the bonds are obtained from reported neutron diffraction data at 300 K \cite{Shiferroelectriclikestructuraltransition2013a} to be as follows. Note the primitive vectors are (units in \AA): $\textbf{a}=(5.06,0,0)$, $\textbf{b}=(-2.53,4.39,0)$, and $\textbf{c}=(0,0,13.2)$.

\begin{center}
\underline{Os-O bonds:}\\
\end{center}
\begin{align*}
\hat{\textbf{b}}_{1}&=(-0.482,-0.668,0.566),\\ 
\hat{\textbf{b}}_{2}&=(0.820,-0.0832,0.566),\\ 
\hat{\textbf{b}}_{3}&=(-0.338,0.752,0.566),\\ 
\hat{\textbf{b}}_{4}&=(0.482,0.668,-0.566),\\ 
\hat{\textbf{b}}_{5}&=(-0.820,0.0832,-0.566),\\ 
\hat{\textbf{b}}_{6}&=(0.338,-0.752,-0.566).\\
\end{align*}
\begin{center}
\underline{Long Li-O bonds:}\\
\end{center}
\begin{align*}
\hat{\textbf{b}}_{7}&=(0.344,0.477,0.808),\\ \hat{\textbf{b}}_{8}&=(-0.685,0.0593,0.808),\\ \hat{\textbf{b}}_{9}&=(0.241,-0.537,0.808),\\ \hat{\textbf{b}}_{10}&=(-0.344,-0.477,-0.808),\\ \hat{\textbf{b}}_{11}&=(0.685,-0.0593,-0.808),\\ \hat{\textbf{b}}_{12}&=(-0.241,0.537,-0.808).\\
\end{align*}
\begin{center}
\underline{Short Li-O bonds:}\\
\end{center}
\begin{align*}
\hat{\textbf{b}}_{13}&=(-1,0,0),\\ \hat{\textbf{b}}_{14}&=(0.5,0.866,0),\\ \hat{\textbf{b}}_{15}&=(0.5,-0.866,0).\\
\end{align*}

\begin{figure}[h]
\includegraphics[width=1\columnwidth]{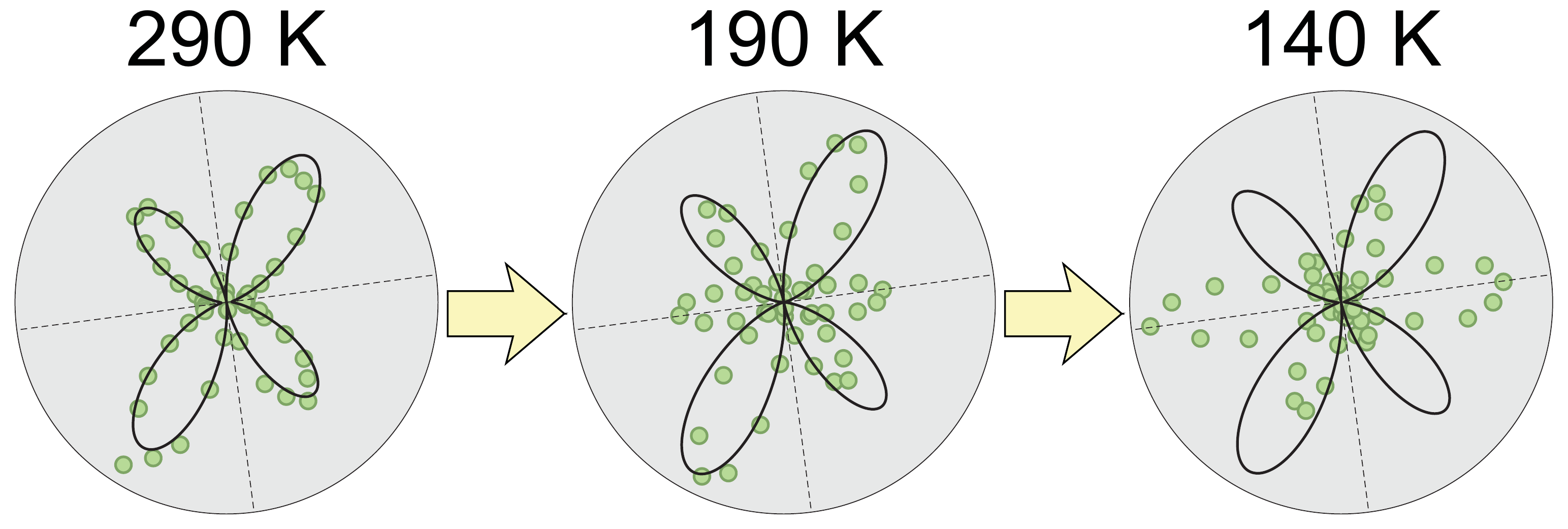}
\caption{\label{angle} Measured SHG-RA patterns (green circles) at 290 K, 190 K, and 140 K. Black curves show patterns calculated using the SHBM, where the $\gamma_{n}^{2\omega}$ values are fixed to the best fit values for the 290 K data, and the $\hat{\textbf{b}}_{n}$'s are varied at each temperature based on neutron diffraction data.}
\end{figure}

To validate our approximation that the $\hat{\textbf{b}}_{n}$ values are constant, we also calculated SHG-RA patterns by fixing $\gamma_{n}^{2\omega}$ at all temperatures and only changing the $\hat{\textbf{b}}_{n}$'s according to lattice parameters reported by neutron diffraction \cite{Shiferroelectriclikestructuraltransition2013a}. As shown in Fig.~\ref{angle}, the calculated SHG-RA patterns hardly change over the interval $\it{T'}>\it{T}>\it{T_{c}}$ and cannot account for our observations.

%